\documentstyle[12pt,fleqn]{article}
\def\beeq{\begin{equation}}
\def\eneq{\end{equation}}
\def\beeqa{\begin{eqnarray}}
\def\eneqa{\end{eqnarray}}

\addtocounter{section}{-1}
\begin{document}

\begin{center}

\mbox{}

\mbox{}

{\Large {\bf Local Non-Fermi Liquid Theory of
Magnetic Impurity Effects in Carbon Nanotubes}}

\mbox{}

\mbox{}

{\large Kikuo Harigaya}

\mbox{}

\mbox{}

{\small {\sl Electrotechnical Laboratory, 
Umezono 1-1-4, Tsukuba 305-8568, Japan}}
\end{center}

\mbox{}

\mbox{}

{\small {\bf 
\noindent
Abstract.}
Magnetic impurity effects in carbon nanotubes are 
studied theoretically.  The multi channel Kondo effect 
is investigated with the band structure of the 
metallic nanotubes.  The local non-Fermi liquid 
behavior is realized at temperatures lower than the Kondo 
temperature $T_K$.  The density of states
of localized electron has a singularity $\sim |\omega|^{1/2}$
which gives rise to a pseudo gap at the Kondo resonance
in low temperatures.  The temperature dependence of the 
electronic resistivity is predicted as $T^{1/2}$, and the
imaginary part of dynamical susceptibilities has the
$|\omega|^{1/2}$ dependence.

\mbox{}
\mbox{}

\begin{center}
{\large {\bf INTRODUCTION}}
\end{center}

\mbox{}

Recently, carbon nanotubes with cylindrical graphite 
structures have been intensively investigated.  Many 
interesting experimental as well as theoretical researches 
have been performed (1,2), and the fundamental metallic 
and semiconducting behaviors of single wall nanotubes
predicted by theories have been clarified in 
tunneling spectroscopy experiments.

In this paper, we will study effects of a magnetic
impurity, for which several experimental works have 
been reported (3,4).  Two channels of electronic 
states are present at the Fermi energy in metallic 
carbon nanotubes.  In the magnetic systems (5), 
the non-Fermi liquid behaviors, i.e., the singular 
density of states and the power law temperature 
dependence of the electric resistivity, have been 
observed experimentally and explained theoretically 
by using the Kondo model or the Anderson model with 
multi channel scatterings.  The similar effects can 
occur in the carbon nanotubes when the presence of 
the two scattering channels plays an important role.

\mbox{}

\mbox{}

\begin{center}
{\large {\bf SPECTRAL FUNCTIONS AND DENSITY OF STATES}}
\end{center}

\mbox{}

The set of integral equations for the multi channel 
Kondo problem can be solved analytically in the limit 
of low frequency and low temperatures (6,7).  We look
at the spectral functions for the empty states at 
$\omega > E_0$, $A_{d,b}^{(+)} (\omega)$, and those for 
the occupied states at $\omega < E_0$, $A_{d,b}^{(-)} 
(\omega)$, where $E_0$ is the ground state energy of 
the magnetic impurity at zero temperature.

After calculations following Ref. (6), we find
the following formula at low frequency:
$A_d^{(\pm)} \sim | \Theta (\omega) |^{-\gamma}$
and
$A_b^{(\pm)} \sim | \Theta (\omega) |^{-1}$,
where $\Theta (\omega) \equiv [ (1 + \gamma)/\gamma \times
(E_0 - \omega)/T_K]^{1/(1+\gamma)}$
with $\gamma = M/N$, $M$ is the scattering channel 
number, $N$ is the spin degeneracy, and $T_K$ is the 
Kondo temperature: 
$T_K = D (\gamma \tilde{\Gamma}/\pi D)^\gamma
{\rm exp} (\pi E_d/\tilde{\Gamma})$.
Here, $E_d$ is the localized level energy,
$\tilde{\Gamma}$ is its broadening, and
$D$ is the band cutoff.

For the special case of interests of metallic nanotubes, 
$M=2$ and $N=2$.  Therefore, we find the singular frequency 
dependence around the ground state energy $E_0$:
\beeq
A_d^{(\pm)} (\omega) \sim A_b^{(\pm)} (\omega)
\sim | E_0 - \omega |^\frac{1}{2}.
\eneq

Next, the density of states of the localized electron 
is calculated by the convolution of spectral functions 
$A_{d,b}^{(\pm)} (\omega)$ (6,7).  The explicit form of 
the density of states for spin $\sigma$ and the channel 
$\alpha$ at $T=0$ becomes $\rho_{\sigma,\alpha} (\omega,0) 
\simeq [ \pi N \tilde{\Gamma} / (1+\gamma)^2]
[1 + \theta(\omega) f_+ (\tilde{\omega})
+ \theta(-\omega) f_- (\tilde{\omega}) ]$,
where
$f_\pm (\tilde{\omega}) = a_\pm |\tilde{\omega}|^{\Delta_{\rm sp}}
+ b_\pm |\tilde{\omega}|^{\Delta_{\rm ch}}$.
Here, $\tilde{\omega} \equiv [(1+\gamma)/\gamma](\omega/T_K)$,
and $a_\pm$ and $b_\pm$ are functions of the scaling 
dimensions of spin and channel fields, $\Delta_{\rm sp} 
\equiv 1/(1+\gamma)$ and $\Delta_{\rm ch} \equiv 
\gamma/(1+\gamma)$.  Both quantities determine the degree 
of singularities of electronic density of states and 
physical quantities at low frequencies.

Specially for metallic carbon nanotubes, we know that
$\Delta_{\rm sp} = \Delta_{\rm ch} = 1/2$.  This leads to
the singularity around the Fermi energy $\omega = 0$:
\beeq
\rho (\omega,0) \sim 1 + \theta(\omega) |\omega|^{\frac{1}{2}}
+ \theta(-\omega) |\omega|^{\frac{1}{2}}
\sim \sqrt{|\omega|}.
\eneq
Such the singular functional form implies that a pseudo gap
develops at the top of the Kondo resonance peak which appear
at temperatures much lower than $T_K$.  There appears a dip 
in the density of states at the Fermi energy.  This is the
local non-Fermi liquid behavior discussed in detail in the 
literature (5).  If it is possible to measure the local 
density of states of a metallic atom attached to the carbon 
nanotubes, for example, by scanning tunneling microscope, we 
could observe such the pseudo gap behavior when role of the 
multi channel scatterings is dominant.

\mbox{}

\mbox{}

\begin{center}
{\large {\bf RESISTIVITY AND MAGNETIC SUSCEPTIBILITY}}
\end{center}

\mbox{}

We consider the electric resistivity in low temperatures.
The scattering rate $\tau$ is calculated from the scattering
$t$ matrix:
$\tau_{\sigma,\alpha}(\omega,T)^{-1} =
-2 {\rm Im} t_{\sigma,\alpha}^{(1)}(\omega+i\delta,T)
= 2 \tilde{\Gamma} \rho_{\sigma,\alpha}(\omega,T)/\rho N$,
where $\rho$ is the density of states at the Fermi energy
of the clean nanotube.  The relation with the electronic 
resistivity $\bar{\rho}(T) \sim [ \int d\epsilon 
( - \partial f/\partial \epsilon)
\tau(\epsilon,T)]^{-1}$
gives the low temperature behavior:
$\bar{\rho}(T)/\bar{\rho}(0) \sim
1 - c (T/T_K)^{{\rm min} (\Delta_{\rm sp},\Delta_{\rm ch})}
+ ... $,
where $c$ is a constant, but it is difficult to obtain
its explicit form only from the information of $\omega$-dependence
of $\rho_{\sigma,\alpha}$.

For the metallic carbon nanotubes, we already know 
$\Delta_{\rm sp} = \Delta_{\rm ch} = 1/2$.  Therefore, 
the low temperature behavior
\beeq
\frac{\bar{\rho}(T)}{\bar{\rho}(0)} \sim
1 - c \sqrt{\frac{T}{T_K}}
\eneq
is expected from the above general formula.

Next, spin and channel susceptibilities are calculated by the
linear response function.  The spin susceptibility is
the magnetic susceptibility in other words.  The imaginary
parts of dynamical susceptibilities are defined as
$\tilde{\chi}''_{\rm sp} = (1/N) {\rm Im} \chi_{\rm sp}$
and
$\tilde{\chi}''_{\rm ch} = (1/M) {\rm Im} \chi_{\rm ch}$.

The first term of $\tilde{\chi}''_{\rm sp}$ at $T=0$ is 
calculated to be
$\tilde{\chi}''_{\rm sp} (\omega,0) \sim
(C_{\rm sp}/T_K) {\rm sgn} \omega 
|\tilde{\omega}|^{(\Delta_{\rm sp}-\Delta_{\rm ch})}$,
where
$C_{\rm sp} = \gamma \Delta_{\rm sp}^2 {\rm sin} (\pi \Delta_{\rm sp})
B(\Delta_{\rm sp},\Delta_{\rm sp})$, and $B(x,y)$ is the 
Beta function.  The second correction gives the $\omega$ 
dependence: $\tilde{\chi}''_{\rm sp} (\omega,0) 
\sim |\tilde{\omega}|^{(2\Delta_{\rm sp}-\Delta_{\rm ch})}$.

In the similar way, the dominant term of 
$\tilde{\chi}''_{\rm ch}$ at $T=0$ becomes
$\tilde{\chi}''_{\rm ch} (\omega,0) \sim (C_{\rm ch}/T_K)\\ 
{\rm sgn} \omega 
|\tilde{\omega}|^{(\Delta_{\rm ch}-\Delta_{\rm sp})}$,
where
$C_{\rm ch} = W_{\rm ch}^2 \Delta_{\rm sp}
{\rm sin} (\pi \Delta_{\rm ch})
B(\Delta_{\rm ch},\Delta_{\rm ch})$.
The second term has the $\omega$ dependence:
$\tilde{\chi}''_{\rm ch} (\omega,0) 
\sim |\tilde{\omega}|^{(2\Delta_{\rm ch}-\Delta_{\rm sp})}$.

The above general formulas reduce to that of 
metallic carbon nanotubes.  The result of singular behavior is 
common for spin and channel susceptibilities:
\beeq
\tilde{\chi}'' (\omega,0) 
\sim A {\rm sgn} \omega (1 - B \sqrt{\frac{|\tilde{\omega}|}{T_K}}
+ ... ),
\eneq
where $A$ and $B$ are constants.  We find $\sqrt{|\omega|}$
dependence at low frequencies.

\mbox{}

\mbox{}

\begin{center}
{\large {\bf SUMMARY}}
\end{center}

\mbox{}

Magnetic impurity effects on metallic carbon nanotubes 
have been investigated theoretically.  We have discussed 
the local non-Fermi liquid behavior at temperatures lower 
than the Kondo temperature $T_K$.  The density of 
states of localized electron has a singularity $\sim 
|\omega|^{1/2}$.  This singular behavior gives rise to a 
pseudo gap at the Kondo resonance in low temperatures.  
The temperature dependence of the electronic resistivity 
is predicted as $T^{1/2}$, and the imaginary part of 
dynamical susceptibilities has the $|\omega|^{1/2}$ 
dependence.

\mbox{}

\mbox{}

\begin{center}
{\large {\bf REFERENCES}}
\end{center}

\mbox{}

\noindent
1. M. S. Dresselhaus, G. Dresselhaus, and P. C. Eklund,
``Science of Fullerenes and Carbon Nanotubes",
(Academic Press, San Diego, 1996).\\
2. R. Saito, G. Dresselhaus, and M. S. Dresselhaus,
``Physical Properties of Carbon Nanotubes",
(Imperial College Press. London, 1998).\\
3. P. Avouris, in ``Electronic Properties of Novel Materials:
Science and Technology of Molecular Nanostructures",
ed. H. Kuzmany, (AIP, New York, 1999).\\
4. L. Grigorian et al, Phys. Rev. B (in press).\\
5. L. Degiorgi, Rev. Mod. Phys. {\bf 71}, 687 (1999).\\
6. E. M\"{u}ller-Hartmann, Z. Phys. B {\bf 57}, 281 (1984).\\
7. P. Coleman, Phys. Rev. B {\bf 29}, 3035 (1984).\\

\end{document}